# Metal-Semiconductor (Semimetal) Superlattices on a Graphite Sheet with Vacancies

**L. A. Chernozatonskii[a], P. B. Sorokin[a,c], E. É. Belova[a], J. Brüning[b], and A. S. Fedorov[c]**

[a] *Emanuel Institute of Biochemical Physics, Russian Academy of Sciences, ul. Kosygina 4, Moscow, 119991 Russia*
*e-mail: cherno@sky.chph.ras.ru*

[b] *Institute of Mathematics, Humboldt University of Berlin, Berlin, 12489 Germany*

[c] *Kirensky Institute of Physics, Siberian Division, Russian Academy of Sciences, Akademgorodok, Krasnoyarsk, 660049 Russia*

It has been found that periodically closely spaced vacancies on a graphite sheet cause a significant rearrangement of its electronic spectrum: metallic waveguides with a high density of states near the Fermi level are formed along the vacancy lines. In the direction perpendicular to these lines, the spectrum exhibits a semimetal or semiconductor character with a gap where a vacancy miniband is degenerated into impurity levels.

PACS numbers: 61.72.Ji, 68.65.Cd, 71.23.-k, 73.21.Cd, 81.05.Uw

It is well known that a carbon material can be either a dielectric (diamond or $C_{60}$ fullerite) or a semimetal (graphite), whereas it can exhibit the properties of a semiconductor and semimetal with a low density of states in the case of nanotubes. New prospects related to the preparation and characterization of individual graphite sheets (graphenes) have been currently developed in carbon nanotechnology [1-3]. Indeed, graphene exhibits an unusual energy spectrum with $\pi$ and $\pi$* bands where the entire Fermi surface is degenerated into points at the intersection of band cones [4]. Therefore, it is a semimetal with a very small amount of free current carriers, in which the energy is proportional to the momentum rather than its square. This special feature of a graphite sheet plays an important role in the electronic structure of carbon nanotubes whose conductivity type depends on the coincidence of the allowed wave vector with the Dirac points of graphene. Recently [5-7], graphene with single vacancies has been theoretically studied. It has been found that a high peak formed by valence electrons localized on a defect appeared in the energy spectrum on the Fermi level [7]. A vacancy reduces the symmetry of the hexagonal lattice of C atoms and removes the degeneracy of the spectrum at the Dirac point. Moreover, a similar situation was observed in graphene with a chain of boundary defects [6], as well as in the electronic structure of carbon nanotubes doped with nitrogen at vacancies [8]. Lehtinen et al. [9] considered the possibility of forming defects of this kind on a graphite surface upon bombardment with helium and hydrogen ions [10]. Lee et al. [11] studied the stability of the spatial arrangement of vacancies in graphene in the framework of the molecular dynamics method. They found that defects of this kind are very stable, and they begin to migrate and combine only at about 3000 K. Actually, this temperature is lower (~1000 K), but is also much higher than room temperature.

In this work, we predict a considerable transformation in the spectrum of graphene in the presence of periodically arranged lines of vacancies at a distance of ~1 nm from each other. The electrons localized on vacancies are combined, and the overlapping wavefunctions of these electrons form a miniband with a high density of states near the Fermi level: the graphene semimetal becomes lined into quasi-one-dimensional metallic nanowaveguides with a high density of carriers alternating with less conductive bands.

## COMPUTATIONAL APPROACH AND CALCULATION DETAILS

Usually, single or double graphite layers are prepared on substrates [1-3]. Therefore, we chose models in which the upper layer with vacancies was molecularly bound to the lower graphite layer. The Abel-Tersoff-Brenner molecular dynamics method (parametrization I) [12], which proved adequate in the calculation of carbon nanostructures [13, 14], was used for structure optimization. The molecular dynamics method was chosen based on consideration for van der Waals forces, which play an important role in the interaction between neighboring graphene sheets. This consideration cannot be taken into account in the DFT-LDA method (based on the density functional theory within the framework of the local-density approximation), which is commonly used for the calculation of the electronic characteristics of carbon nanostructures. The molecular

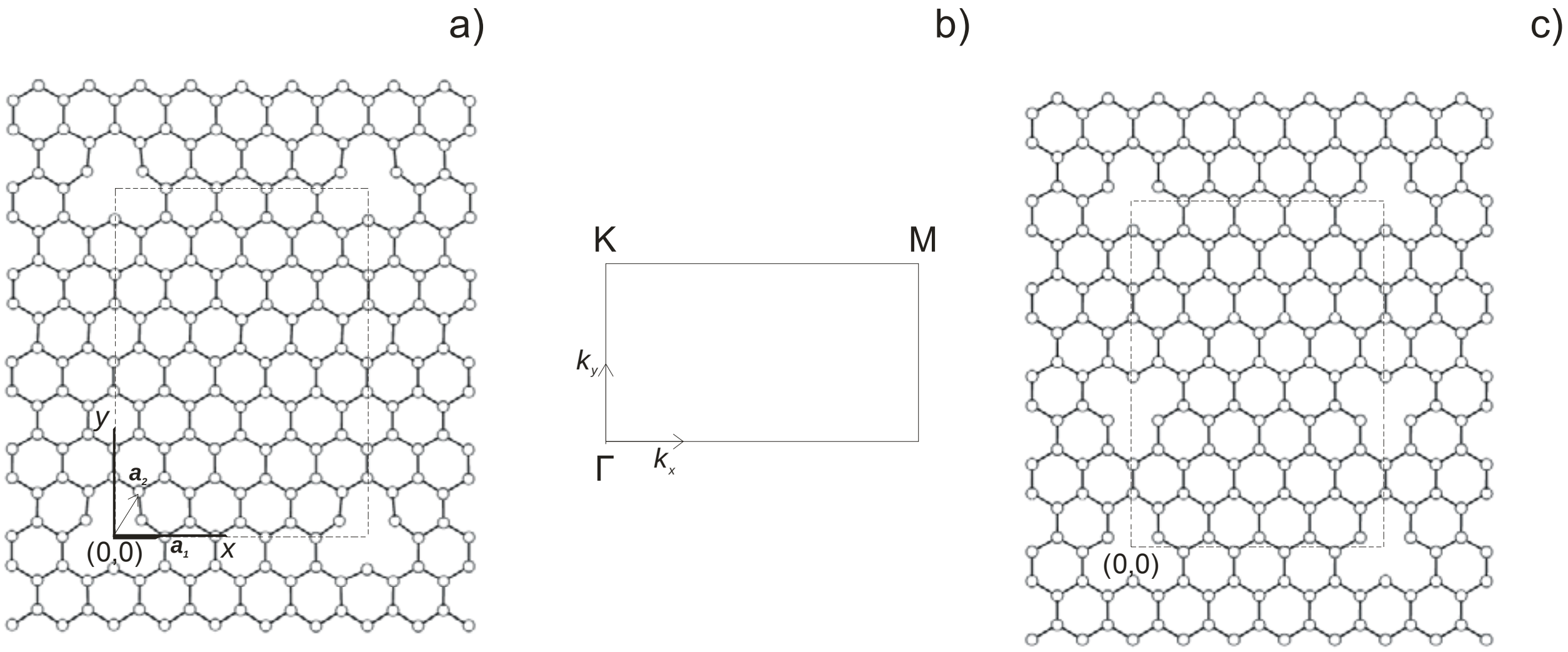

Fig. 1. Graphite planes with periodically arranged vacancies. (a) "Simple" structure *(5,0)-(-4,4)* with a rectangular unit cell (framed). (b) First Brillouin zone for the chosen superlattice; $a_1$ and $a_2$ are the unit cell vectors of graphene. (c) View of a "complex" structure with the same rectangular unit cell containing an additional vacancy at the site (-5/3, 10/3).

interaction potential was chosen in the standard 6-12 form [14].

The computation of the band structure and the densities of electronic states was performed using the OpenMX v. 2.3 program [15] within the framework of the local density functional [16-18].

A linear combination of localized pseudoatomic orbitals was used as a basis [19, 20]. The pseudopotential generated by the Trouillier-Martins method [21] with partial core correction [22] was chosen as a pseudopotential for carbon. The *s2p2d1* set, which was obtained upon the optimization of the *s3p3d2* basis set for a defect-free graphene sheet, was chosen as valence orbitals. The cutoff radius of 4.5 au was chosen for orbitals. In the numerical integration of the Poisson equation, a cutoff energy of 150 Ry was chosen.

To obtain a band pattern, 50 *k* points were used in each of the high-symmetry directions. To calculate the density of electronic states, a 16 x 32 x 1 set of k points was used.

## RESULTS AND DISCUSSION

We studied structures with various arrangements of defects on a graphite sheet. To classify superlattices with a simple unit cell containing only one vacancy, we chose a set of four indices *(n, l)-(m, k),* which denote the vectors of this superlattice in terms of the unit vectors $a_1$ and $a_2$ of the graphite lattice with the center (0,0) at the site of a central vacancy (an analog of the classification of carbon nanotubes [4]). However, to classify complex structures whose cell contains other nonequivalent vacancies, the designations of the location of these latter can contain the indices *n, l'*, *m* and *k'* that are fractional numbers multiple of 1/3 along with integers. In this case, "simple" structures have no center of inversion (for example, in these rectangular superlattices, there is no inversion with respect to the *y* axis (Fig. 1a); this leads to the appearance of polarization (an analog of the 3D piezoelectric effect) on the propagation of phonons in this direction). "Complex" structures can have a center of inversion in the case of rectangular superlattices, as exemplified in Fig. 1c.

It is extremely important that, as in the case of carbon nanotubes, the electronic structure of a graphite sheet is changed under changes of the configuration of vacancies. Thus, relatively closely spaced defects can form an additional band near the Fermi energy, as in the case of an individual fragment of a graphite sheet with hydrogen atoms added at the ends [23].

The width and shape of the band under consideration entirely depend on the mutual arrangement of defects. It is also reasonable to expect the appearance of van Hove singularities because of the presence of a superlattice, which is an analog of unrolled nanotube strips [4]. Therefore, as in nanotubes, each particular vacancy structure corresponds to an intrinsic set of peaks in the electronic density of states (DOS) as a fingerprint. This fingerprint manifests itself in all optical spectroscopic experiments (Raman spectra, luminescence, fluorescence, and resonance optical effects), as well as in the studies of single-layered carbon nanotubes [4].

For simplicity, we considered the structures of rectangular vacancy rows where the pairs *(n, l)* and *(m, k)* determine the periods of vacancies in the *x* and *y* directions, respectively, and the Brillouin zone of which is also rectangular (see notations in Fig. 1). As can be seen in Fig. 1, in a defect graphite structure with the indices (5,0)-(-4,4), vacancies are closest to each other along the *x* direction. Therefore, its band structure (Fig. 2a) forms a conduction miniband in the *K-M* direction with the width $\delta E_c$ = 0.29 eV and a high density of electronic states near the energy $E = 0,$ which corresponds to the Fermi level $E_F$ (cf. Fig. 2b). It is separated from the first unoccupied band by the gap $A_v$ = 0.23 eV and from the nearest occupied band by the gap $A_c$ = 0.23 eV (a similar result was obtained for the structure shown in Fig. 1c).

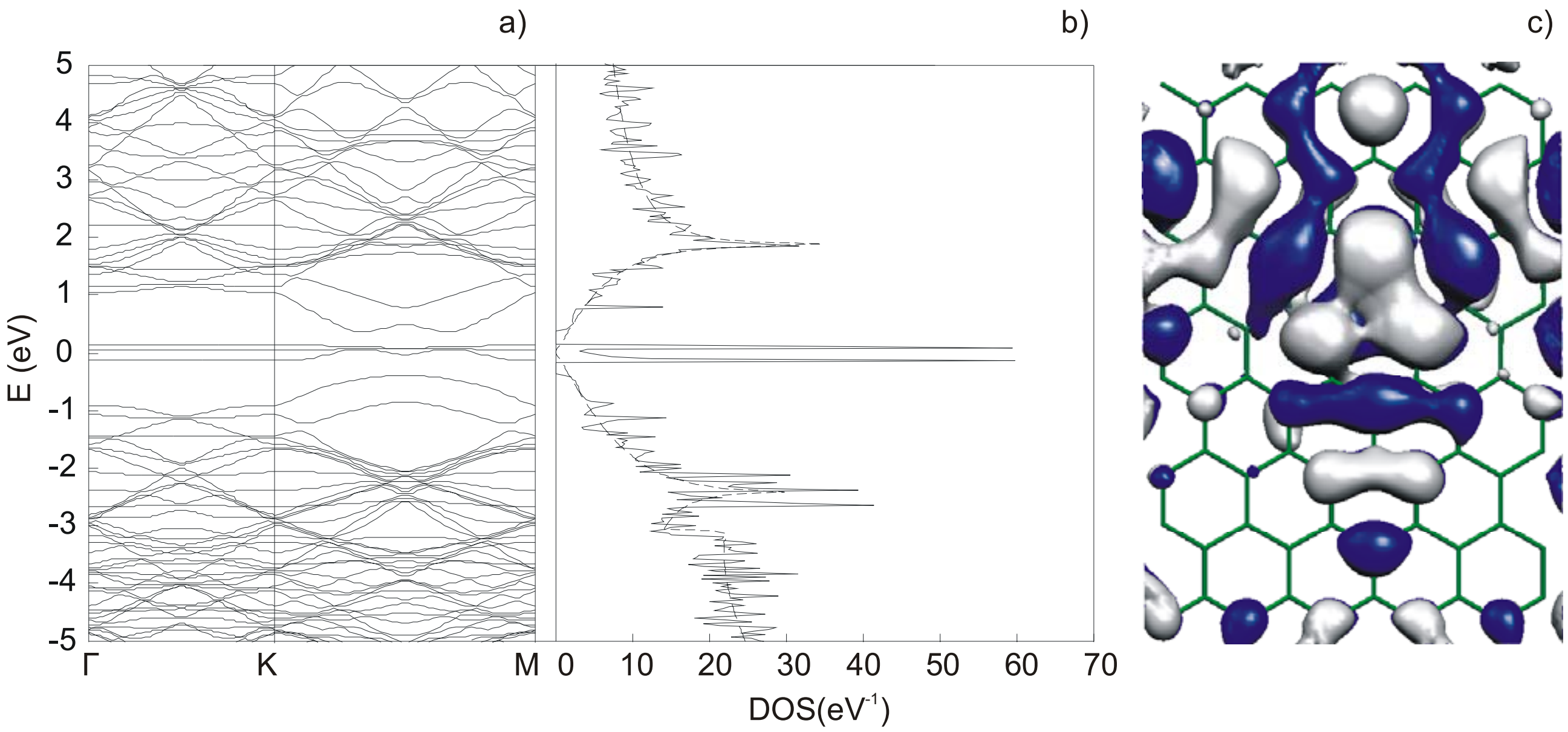

Fig. 2. (a) Band structure and (b) density of electronic states for (solid line) the structure (5,0)-(-4,4) and (dashed line) a defect-free graphite sheet. The Fermi energy was taken as zero. Fig. 1a shows the view of the structure. (c) HOMO-level orbitals at the *k* point (0,0.5) of the Brillouin zone (cutoff value of 0.01): dark and bright figures correspond to unlike signs of the wavefunction.

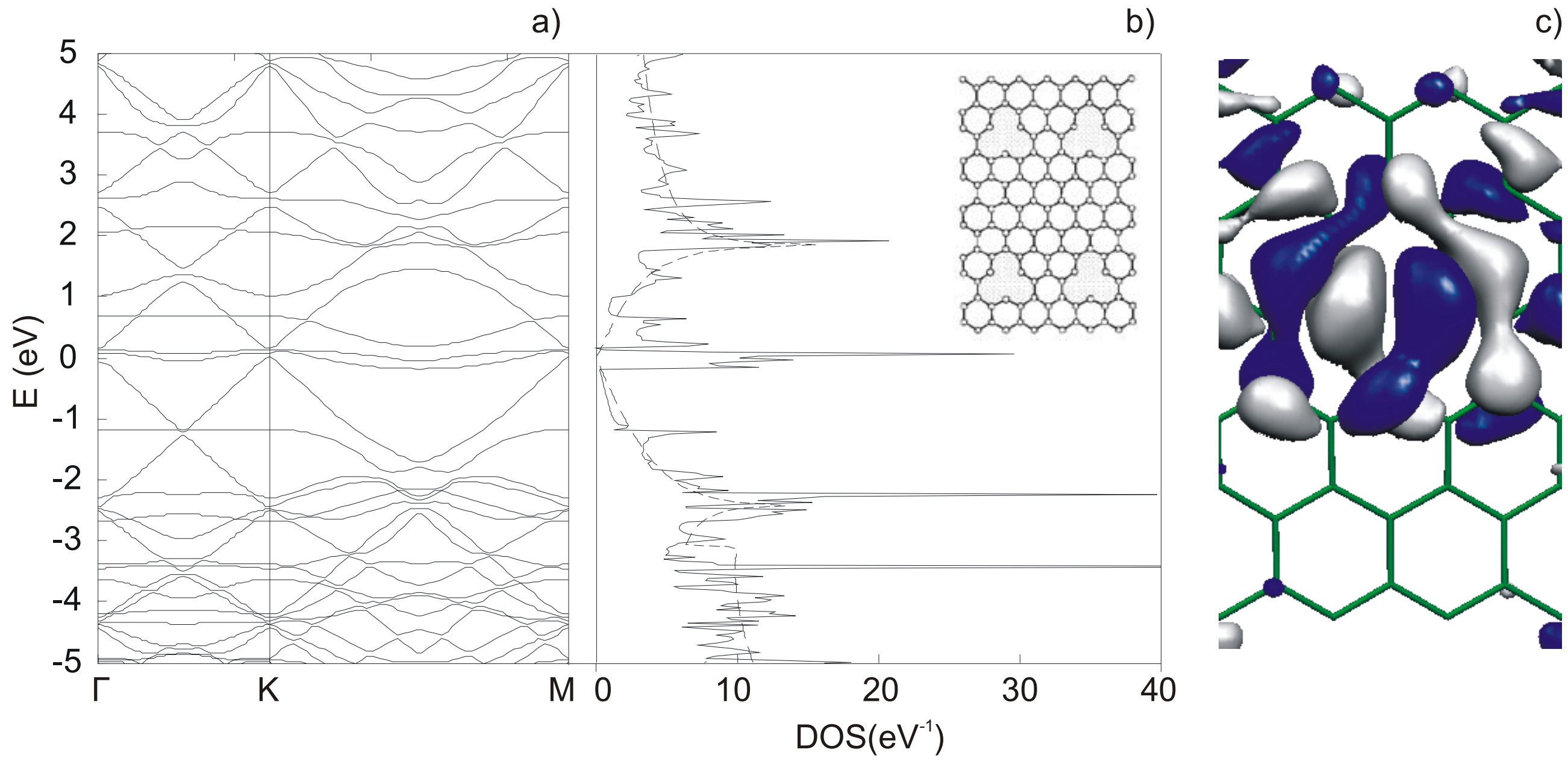

Fig. 3. (a) Band structure and (b) density of electronic states for (solid line) the structure (3,0)-(-3,3) and (dashed line) a defect-free graphite sheet. The insert shows the view of the structure. The Fermi energy was taken as zero. (c) HOMO-level orbitals at the *k* point (0.5, 0) of the Brillouin zone (cutoff value of 0.01): dark and bright figures correspond to unlike signs of the wavefunction.

Fig. 2c shows the surface of orbitals at the *k* point (0,0.5), which characterize electron density at the Fermi level. At this *k* point, the Fermi level coincides with the high occupation molecular orbital (HOMO) level. It can be seen that electron density at $E = E_F$ is concentrated at the "line" of vacancies (with the width 1 nm = $6d_{c\text{-}c}$ at the period $D_y = 12d_{c\text{-}c} \sim 2$ nm; $d_{c\text{-}c} \sim$ 0.145 nm is the distance between neighboring C atoms), and strips between the "lines" are depleted in current carriers. This is confirmed by the shape of the spectrum in the *Γ-K* direction, where the repulsion of the spectrum of defect-free graphene occurs at the Fermi energy; therefore, a local band gap of width 1.9 eV is formed in this direction. The van Hove singularities of this band gap manifest themselves as corresponding great peaks in the plot of the DOS(E) function. In this gap, impurity levels corresponding to electrons localized at "lines" are arranged, as is the case in the localization of electrons in a superlattice with quantum wells [24]. Thus, this structure can be characterized as a two-dimensional superlattice of alternating metallic "lines" and semiconducting nanostrips. Closely spaced vacancies, for example, as in the structure (3,0)-(-3,3) in Fig. 3, lead to the interaction of electrons localized on them in both directions. In the spectrum (Fig. 3a), a "metallic" miniband appears in both the *Γ-K* direction (a narrower of 0.1 eV) and the *K-M* direction (a broader of 0.2 eV), although the degeneracy at the points of intersection of $\pi$ and $\pi$* bands *Γ, K,* and M is removed. The partial localization of the orbitals of vacancies most distant from each other along the $x$ direction is retained; this manifests itself in the appearance of two levels in the $\Gamma$-$K$ region. They and their extension to the $K$-$M$ region contribute to strong van Hove peaks in DOS at energies of 0.69 and -1.18 eV (Fig. 3b). The DOS(E) function shows that, in spite of its peak character, no band gaps appear in the spectrum and, on the average, it reproduces the relation for a defect-free graphite sheet. It can be seen in Fig. 3c that, in this case, the electron density at $E = E_F$ is concentrated at vacancy "lines" of width ~1 nm, whereas sites between the "lines" are depleted.

Based on Fig. 2 and Fig. 3, we can conclude that the behavior of bands in the spectra of *(n, l) - (m, k)* vacancy structures obeys a simple law: if the index difference $|n - l|$ $(|m - k|)$ is not a number divisible by 3, a band gap appears in the $\Gamma$-$K$ $(M$-$K)$ direction (similarly to the case of an *(n, m)* carbon nanotube, which exhibits metallic properties if the difference of its indices is a multiple of 3 or semiconducting properties in any other case [4]).

It is reasonable to expect that an individual vacancy "line," which can be considered as a metallic nanowaveguide on a semimetallic graphene plane, will exhibit a high degree of metallicity. In this context, the question arises: How it will manifest itself in its physical properties? The high conductivity of "lines" and DOS peaks will be detected in measurements by scanning tunneling microscopy and spectroscopy (for example, cf. [24, 25]). The same waveguide can serve as a spin waveguide because an unpaired spin remains on a vacancy and vacancies can result in the formation of magnetic moments over a considerably large area [5, 6, 9]. For example, the interaction of these spins on vacancies even at sufficiently large distances of ~1 nm can explain magnetism in purely carbon systems, for example, in polymerized fullerites [26].

It is most likely that magnetism can be suppressed by doping with nitrogen at vacancies, as considered for nanotubes [8], without changing considerably the degree of metallicity of vacancy nanowaveguides. Then, a phase transition to a superconducting state can occur in the above nanowaveguides at low temperatures, and a superlattice of superconducting nanostrips can be obtained on graphene. It is most likely that a change in the phonon spectrum, which exhibits the waveguide properties of the structure, will influence thermal and electron-phonon effects in the vacancy superlattice on graphene. We will consider these situations in the future.

Thus, in this work, we found that, in the development of a structure of periodically closely spaced monatomic vacancies on a graphite sheet, a self-doping effect occurs in vacancy-free graphene: a charge redistribution between lattice atoms and defects, in which quasi-homogeneous vacancy "lines" become metallic with a high density of carriers.

For example, such vacancy superlattices and nanowaveguides on graphene will be produced by knocking out carbon atoms point by point using the tip of scanning tunneling microscope as an electron-beam nanosource (by the way, a tip of a carbon nanotube can play this role [4]). We found that the use of defect "lined" metallic structures based on graphene in new electronic and spintronic nanodevices is promising.

We are grateful to the Joint Supercomputer Center of the Russian Academy of Sciences for the use of a cluster computer for quantum-chemical calculations. The molecular orbitals were visualized using the Molekel 4.0 program [27]. The geometry visualization was performed using the ChemCraft program [28]. This work was supported by the Russian Foundation for Basic Research (project no. 05-02-17443) and Deutsche Forschungsgemeinschaft/Russian Academy of Sciences (DFG/RAS, project no. 436 RUS 113/785).